\documentclass[twoside]{article}

\usepackage{PRIMEarxiv}

\usepackage[utf8]{inputenc} 
\usepackage[T1]{fontenc}    
\usepackage{hyperref}       
\usepackage{url}            
\usepackage{booktabs}       
\usepackage{amsfonts}       
\usepackage{nicefrac}       
\usepackage{microtype}      
\usepackage{subcaption}
\usepackage{lipsum}
\usepackage{fancyhdr}       
\usepackage{graphicx}       
\usepackage{siunitx}
\graphicspath{{media/}}     
\usepackage{cancel}
\usepackage{multicol}
\usepackage{hyperref}
\usepackage{booktabs}
\usepackage[utf8]{inputenc}
\usepackage[cmex10]{amsmath}
\usepackage{algorithm}
\usepackage{algorithmic}
\renewcommand{\figurename}{Fig.}
\renewcommand{\tablename}{Table}

\newcommand{\figref}[1]{\textcolor{blue}{\figurename~\ref{#1}}}
\newcommand{\tabref}[1]{\textcolor{blue}{\tablename~\ref{#1}}}

\hypersetup{
    colorlinks=true,
    linkcolor=blue,
    citecolor=blue,
    filecolor=blue,
    urlcolor=blue
}

\usepackage{multirow}

\title{Enhancing Offshore Wind Simulations: Interpolated
Switching via DLL Black-Boxes
\thanks{\textit{\underline{\textbf{In Review} at IET Renewable Power Generation}} \\} 
}

\author{Nicolae Darii,   Ranjan Sharma, Vladislav Akhmatov\\  Kanakesh Vatta kkuni, Chi Su, Oscar Sabor\'{i}o-Romano, Nicolaos A. Cutululis}

\begin{document}
\maketitle

\begin{abstract}
The modern power system, increasingly composed of Inverter-Based Resources (IBR) from multiple manufacturers, requires new study and design techniques that balance accuracy with the need to protect the Intellectual Property (IP) of various stakeholders. One possible method to support detailed electromagnetic transient (EMT) simulations is to convert the original equipment manufacturers' (OEM) models into shareable black-box versions using dynamic link libraries (DLLs). This technique prevents IP violations while potentially maintaining simulation accuracy by embedding the original components within the shareable DLL. Thereby, this work aims explicitly to enhance simulation fidelity by translating full-switching models of offshore wind turbines (OWTs). In this context, the paper offers valuable recommendations, including how to convert interpolation-based elements, preserve simulation speed, recognize limitations, and outline future improvements.
\end{abstract}


\section{Introduction}
The increasing share of renewable energy sources (RES) and their inverter-based resources (IBRs) brings, along with net-zero
benefits, as well as new challenges regarding interoperability \cite{Gharehpetian2024FutureSolutions}. Bringing in more IBRs will require multiple players to operate within a context of intellectual property (IP) protection \cite{Jahn2022AnProtection, 2025DefinitionPUBLIC}, which cannot be allowed to impede the need for a system that functions correctly. 
This significantly distinguishes the upcoming power system from the classic one, as the former devices were more uniform and standardized in modeling terms and easier to label in terms of parameters by just knowing their nominal values and controller types. On the other hand, the new IBR-based devices can be more vendor-specific and characterized by internal structures that are distinctly different from one another.
In addition, it is necessary to model and use tools such as Electromagnetic Transients (EMT) simulations to capture faster phenomena accurately \cite{Deng2024AStudy}, which are primarily caused by IBR's controls and the switching of IGBT-based devices. The inclusion of such phenomena can be achieved by accurately modeling devices in EMT-based software.

When studies are conducted for high-level estimation, generic models can usually be implemented; however, analyzing critical phenomena such as fault reactions or device interactions requires more detailed models \cite{CSEECIGRE}. Unfortunately, sharing such detailed models is often impracticable due to intellectual property protection — a \textit{sine qua non} for maintaining competitiveness among power system stakeholders.

Initially introduced by IEC \cite{IECIEC} and expanded by CIGRE \cite{CSEECIGRE}, the concept involves using Dynamic Link Libraries (DLL) based components to incorporate the most critical parts of the models, typically represented by the control structure in firmware form. These parts are often the same ones used by the Original Equipment Manufacturers (OEM). The application of DLLs for EMT model sharing is a niche area with limited general literature, since it primarily targets industrial users and applications with encrypted methods and models.

Additionally, to improve the accuracy of the DLL versions of the shared models, it is necessary to include switching models \cite{Beddard2016ImprovedConverters}. However, to do so, the generated DLL must be compatible with interpolation features, which are distinctive of switching-based devices. This is necessary to accurately estimate the switching instant in a context of high switching frequency and a fixed time step \cite{Manitoba-HVDCResearchCentre2010EMTDC-TransientSimulation}. Additionally, its implementation should not compromise the simulation's speed. Therefore, the paper addresses the DLL conversion with switching models by demonstrating how to convert the interpolated elements (firing pulse generators and logic ports) into their DLL equivalents. Thus, testing the accuracy of the switching-based DLL by comparing its results with those of the original EMT model, and assessing its efficiency by comparing total CPU time. The validation process is conducted in steps: initially, verifying the base elements, then extending the study to the entire benchmark in both regular and fault operation, culminating in a validation conclusion through harmonic comparison.

The research summary results show that the switching models implemented via DLL are practical and consistent with the original version, while also advancing empirical methods to avoid slow DLLs. Limitations regarding the use of DLL for time steps other than the intended one have been identified. Additionally, inherent delays in the EMT software may introduce bias in negative scenarios. Future efforts should focus on improving the accuracy and speed of the DLL-based models.

\section{Workflow}
To enhance the black-box fidelity of offshore wind turbines, it is necessary to encapsulate the original OEM models into a DLL compatible with commercial EMT-based software. Additionally, to incorporate fast phenomena, switching models must be incorporated into the black-box models. However, commercial EMT software is generally based on fixed-time-step simulation; therefore, it includes interpolation features and devices that can address issues related to phenomena that occur within a time step \cite{Marthi2023InterpolationInverters}. Eventually, the DLL conversion should not slow down simulations, otherwise resulting in a \textit{Ultra-Solution} \cite{Watzlawick1988Ultra-Solutions:Successfully}, situation in which persistent efforts to solve or optimize a problem, such as the DLL model conversion, not only fail but ultimately worsen the original condition, such as having to deal with model inaccuracies or slowdowns originating from the DLL. Therefore, in this section, some empirical indications concerning the DLL speed will also be presented. 
 
Although it is highly recommended to use this procedure primarily for code-based logics that are desirable to keep confidential, the DLL conversion can also be implemented on control structures that are not strictly implemented in the final firmware but are still present in the OEM EMT models. To do so, it is possible to utilize software that includes Auto-Gen Code (automatic generation), such as MATLAB/Simulink. 
Therefore, parts of the original model that are included directly as code in the final hardware are directly convertible into a DLL. 

\subsection{Inclusion of interpolation features in DLL}
\begin{figure}
    \centering
    \includegraphics[width=0.8\linewidth]{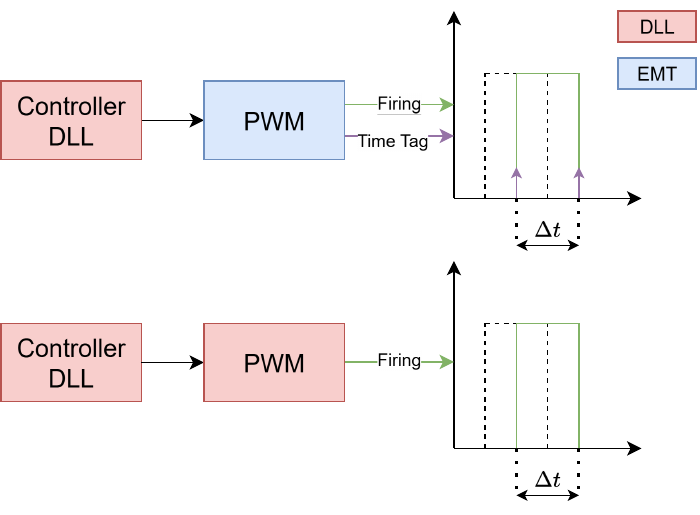}
    \caption{Difference in implementing switching in DLL and software}
    \label{fig:difference}
\end{figure}

In principle, studies that incorporate switching behaviors into models are more accurate than average models \cite{Deng2024AStudy}. Typically, the DLL would involve just the firmware part; thus, the DLL would output the voltage reference signals later used by the switching EMT module. However, introducing the switching module (which may involve additional firing logics that should be IP-protected) in the DLL would generate signals used directly by the EMT IGBT modules. In this case, the generation of relatively high-frequency firing signals, on the order of kHz (depending on the OEM), will produce outputs that occur initially within a single timestep, as shown in \figref{fig:difference}. In principle, EMT software, such as PSCAD, includes interpolation-based components that allow for the accuracy to be maintained even during these types of occurrences \cite{Marthi2023InterpolationInverters, Manitoba-HVDCResearchCentre2010EMTDC-TransientSimulation}.

To incorporate the switching behavior into EMT software using the Interpolation feature, it is essential to embed it within the DLL. Specifically, the interpolation for firing signal generators should include estimating the exact crossing time, even if it occurs within a single time step. 

\begin{figure}
    \centering
    \includegraphics[width=0.8\linewidth]{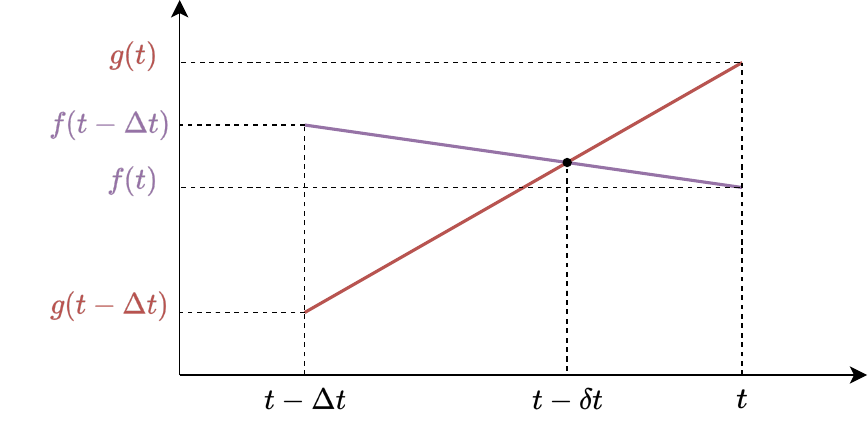}
    \caption{Interpolation}
    \label{fig:interp}
\end{figure}

The interpolation mechanism, introduced in \cite{Manitoba-HVDCResearchCentre2010EMTDC-TransientSimulation}, is analytically explicated in \eqref{enq: interp} and visually represented in \figref{fig:interp}.

\begin{equation}
    \delta t = \frac{f(t)-g(t)}{f(t)-f(t-\Delta t)-g(t)+g(t-\Delta t)} \Delta t
    \label{enq: interp}
\end{equation}

As seen, it is possible to express the interpolation delay $\delta t$ by weighting the simulation time step $\Delta t$ with a combination that depends just on the values of the carrier $f(t)$ and modulant $g(t)$ at the current time step, and the ones saved as states from the previous $f(t-\Delta t)$ and $g(t-\Delta t)$. This approximation holds until the timestep $\Delta t$ is small enough to have an approximately linear behavior around the crossing point $t-\delta t$.
Before introducing the further logic, the Heaviside function in \eqref{eqn: heavy} is defined hereafter that will be used in the further formulation to indicate a function that becomes one once its argument is greater than 0.

\begin{equation}
\mathcal{H}(x) = 
\begin{cases}
1 & \text{if } x > 0 \\
0 & \text{otherwise}
\end{cases}
\label{eqn: heavy}
\end{equation}

Since the formulation in \eqref{enq: interp} would be calculated at any time step, regardless of whether there is a crossing of the signals, it is necessary to implement an additional layer that enables the calculation only when signals have crossed. Therefore, by analytically expressing the logic, the Heaviside function, used on the difference of the signals, is used as a detector for the cases where the modulant and the carrier cross, and two cases are identified as \eqref{eqn: cases}.

\begin{equation}
A = \mathcal{H}(f(t) - g(t)), \quad B = \mathcal{H}(f(t-\Delta t) - g(t-\Delta t))
\label{eqn: cases}
\end{equation}

Calculating interpolation when it is not essential is resource consuming. It is optimal to enable the logic only when it is required. The final $\text{Time-Tag}$ is, thus, a combination of the delay $\delta t$ computed through \eqref{enq: interp} and a \textit{XOR} logic on the modulant-carrier values as shown in \eqref{eqn:tt}.

\begin{equation}
    \text{$\text{Time-Tag}$} = \delta t\cdot (A \oplus B)
    \label{eqn:tt}
\end{equation}
\subsection{Elements with interpolation feature}
In addition to the interpolation firing signal, which takes as input the four values to generate an output tuple composed of, on one side, the boolean firing signal and, on the other, the scalar $\text{Time-Tag}$ defined as \eqref{eqn: firing}.

\begin{equation}
f : \mathbb{R}^4 \longrightarrow \{ \text{1}, \text{0} \} \times \mathbb{R}
\label{eqn: firing}
\end{equation}

Moreover, there are also other elements, such as interpolated logic ports \cite{Manitoba-HVDCResearchCentre2010EMTDC-TransientSimulation}, that require a correct translation. The overall switching logic includes \textit{NOT} and \textit{AND} ports to generate the firing signal simultaneously for the inferior IGBT. 

\subsubsection{Interpolated \textit{NOT} port}
\begin{figure}[h]
    \centering
    \includegraphics[width=0.5\linewidth]{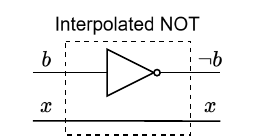}
    \caption{Interpolated NOT}
    \label{fig:NOT}
\end{figure}

The interpolated \textit{NOT} port has a straightforward translation where just the boolean firing signal $b$ undergoes transformation and comes out simply as its negated form $\neg b$. The scalar $\text{Time-Tag}$ $x$ is unaltered as shown in \eqref{eqn: not} and \figref{fig:NOT}.

\begin{equation}
f : (b, x) \in \{ \text{1}, \text{0} \} \times \mathbb{R} \mapsto (\neg b, x) \in \{ \text{1}, \text{0} \} \times \mathbb{R}
\label{eqn: not}
\end{equation}

\subsubsection{Interpolated \textit{AND} port}
\begin{figure}[h]
    \centering
    \includegraphics[width=1\linewidth]{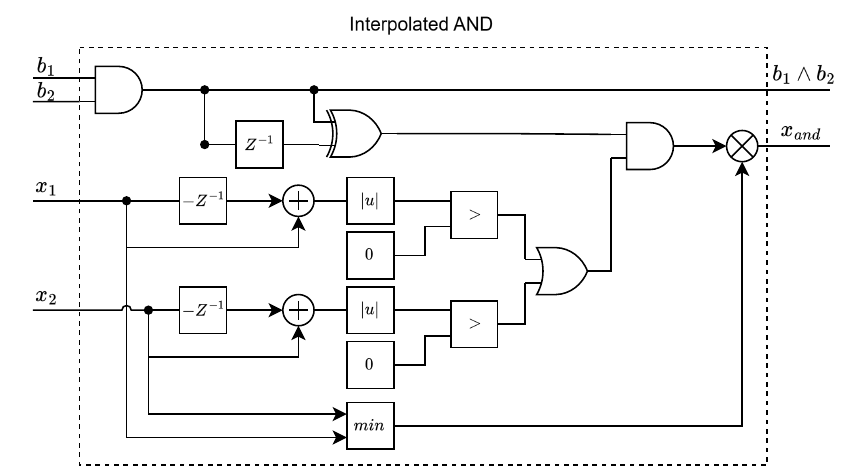}
    \caption{Interpolated AND}
    \label{fig:AND}
\end{figure}

To correctly adapt interpolated \textit{AND} ports, additional logic must be applied to the $\text{Time-Tag}$ signals as showed in \eqref{eqn:and} and \figref{fig:AND}. Without this, the system would emit a $\text{Time-Tag}$ whenever either input tuple changes, instead of only when the final port output changes. To address this, a control logic function $Y$ is introduced into the $\text{Time-Tag}$ computation.

\begin{equation}
\begin{aligned}
f :\ ((\{ \text{1}, \text{0} \} \times \mathbb{R}) \times (\{ \text{1}, \text{0} \} \times \mathbb{R})) 
\rightarrow \{ \text{1}, \text{0} \} \times \mathbb{R} \\
((b_1, x_1), (b_2, x_2)) \mapsto \\(b_1 \land b_2,\ (\min(x_1,x_2))\cdot Y(x_1, x_2, b_1, b_2))
\end{aligned}
\label{eqn:and}
\end{equation}

The function $Y$ acts as a gatekeeper for the $\text{Time-Tag}$ signal and is defined to return 1 only when both input signals experience a simultaneous change in value and $\text{Time-Tag}$, as shown in \eqref{eqn:heavy}.

\begin{equation}
\begin{aligned}
\text{AND}(\cdot) &= \mathcal{H}\Bigl[ \bigl(\Delta(b_1\land b_2)[t] \oplus \Delta(b_1\land b_2)[t-\delta t]\bigr) \\
&\quad\cdot \bigl( \mathcal{H}(|\Delta x_1|) + \mathcal{H}(|\Delta x_2|) - 1 \bigr) \Bigr]
\end{aligned}
\label{eqn:heavy}
\end{equation}

This compact logic ensures that the minimum (since it is desired to use as reference the last switching signal) of both inputs' $\text{Time-Tag}$ values appears only when the final \textit{AND} port output changes due to both inputs, preventing false triggers. The logic can be implemented directly in code or designed graphically in Simulink and translated into C through automatic code generation. The interpolated \textit{AND} maybe used for specific operation such as blocking the PWM in the model, thus in the DLL.

\subsection{Generate a DLL black-box model}
\begin{figure*}[ht!]
    \centering
    \includegraphics[width=0.8\linewidth]{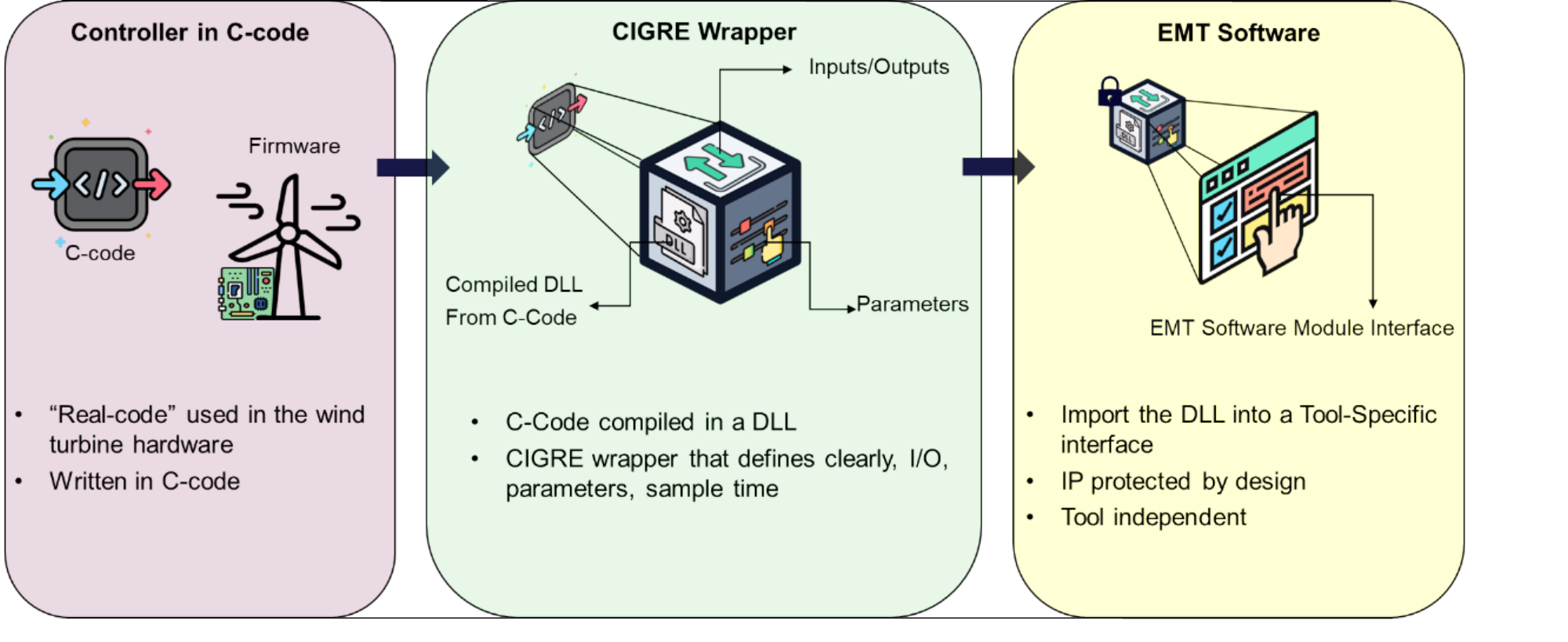}
    \caption{DLL generation and usage scheme}
    \label{fig:dll}
\end{figure*}

The first step is to convert the EMT model into a DLL-compatible version. To do so, there is a proposal of standard procedures initially formulated by IEC \cite{IECIEC} and further explored by CIGRE \cite{GuidelinesECIGRE}.
The general procedure can be decomposed into three phases, where each part is assigned to a different type of user as shown in \figref{fig:dll}:

\begin{enumerate}
    \item Design/Collect protections or control procedures in the form of code, potentially the firmware that is included in the final hardware.
    \item Write/Add a \textit{wrapper} that links inputs, outputs, and parameters of the base firmware. 
    \begin{enumerate}
        \item Static items: model name, version, descriptions, sampling time, number of inputs/outputs
        \item Dynamic items: inputs, outputs, parameters, state, real-time
        \item States variables: designed to be unique in the code, potentially definable for the initial state condition, and useful for snapshot-based software.
    \end{enumerate}
    During the generation of the DLL, it is possible to decide which items are accessible or not from the DLL's final interface for IP protection.
    \item Import the DLL into the final software directly, or use aid methods such as the RTE proposed in \cite{GitHubRte-france/PSCAD-import-tool-for-IEEE-CIGRE-DLLs}.
\end{enumerate}

\subsection{DLL speed-up}
If the original C code is poorly conditioned, it may result in a DLL that, despite being precompiled, performs more slowly than a version built directly within the final EMT software. Therefore, three principal suggestions have been identified to help reduce the DLL's speed bottleneck, aiming to match the performance of a structure built entirely in EMT using in-house elements. The indications, resulting from an iterative study, are:

\begin{enumerate}

    \item Minimize blocks that include states to only the essential minimum. When possible, replace state-containing blocks with basic arithmetic operations and standard mathematical functions, keeping these blocks to a minimum. When optimizing the code to reduce states, verify and confirm that the optimized DLL matches the original DLL in accuracy.
    \item For DLL routines that run only during particular events or not at each time step, structure the code to avoid executing these sections at every step.
    \item In the case that PWM's input signals are pure sine waves $z(t)$, it is possible to generate a delayed signal $z(t)'$ without memory allocation as shown in \eqref{eqn: sindelay}, where $\delta t$ is the delay time and $\omega_0$ is the frequency of the modulant signal. 

    \begin{equation}
    z(t)'=\cos \left(\arcsin(z(t))+\frac{\pi}{2}- \mathcal{H}(z(t))\cdot \omega_0\cdot\delta t\right)
    \label{eqn: sindelay}
    \end{equation}
    
\end{enumerate}

\section{Results}
\begin{figure*}[t]
    \centering
    \begin{subfigure}[b]{0.32\textwidth}
        \includegraphics[width=\linewidth]{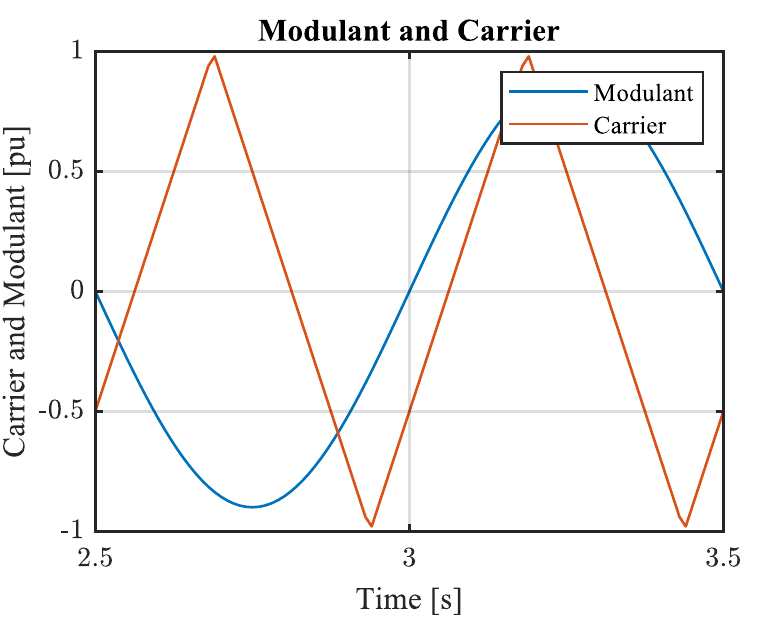}
        \caption{PWM - Large time step}
        \label{fig:pwm-fire}
    \end{subfigure}
    \begin{subfigure}[b]{0.32\textwidth}
        \includegraphics[width=\linewidth]{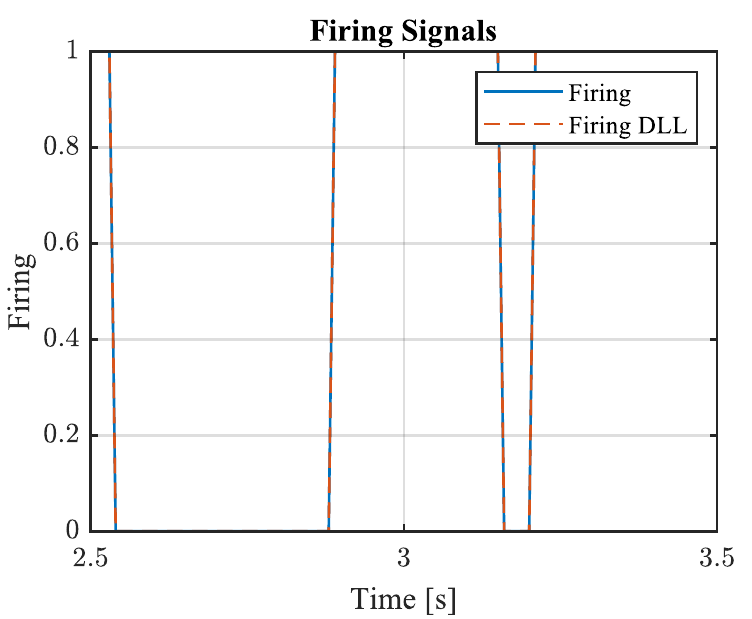}
        \caption{Firing - Large time step}
        \label{fig:fi-1s}
    \end{subfigure}
    \begin{subfigure}[b]{0.32\textwidth}
        \includegraphics[width=\linewidth]{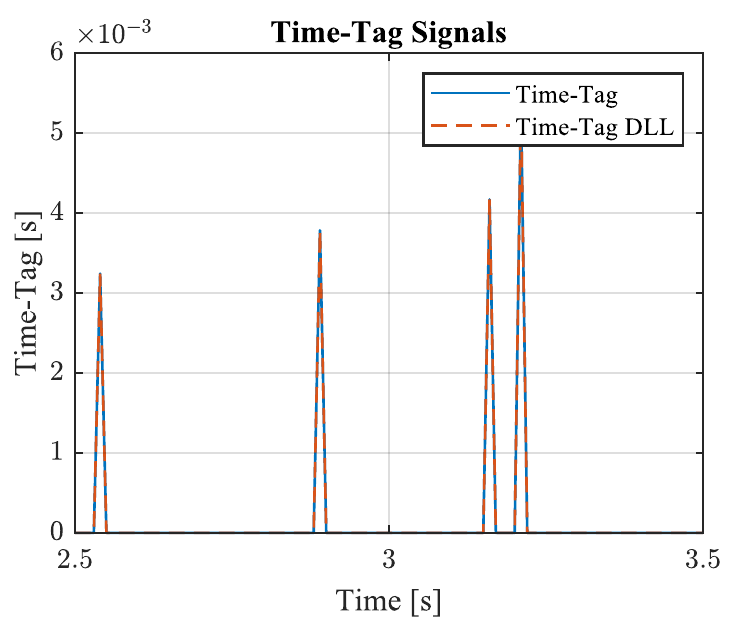}
        \caption{$\text{Time-Tag}$ - Large time step}
        \label{fig:tt-1s}
    \end{subfigure}

    \vspace{0.5em}

    \begin{subfigure}[b]{0.32\textwidth}
        \includegraphics[width=\linewidth]{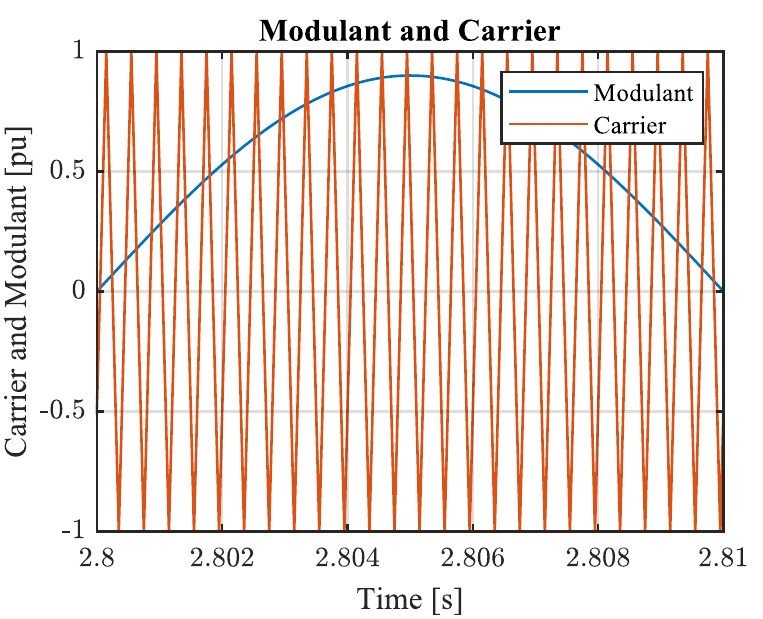}
        \caption{PWM - Small time step}
        \label{fig:pwm-fast}
    \end{subfigure}
    \begin{subfigure}[b]{0.32\textwidth}
        \includegraphics[width=\linewidth]{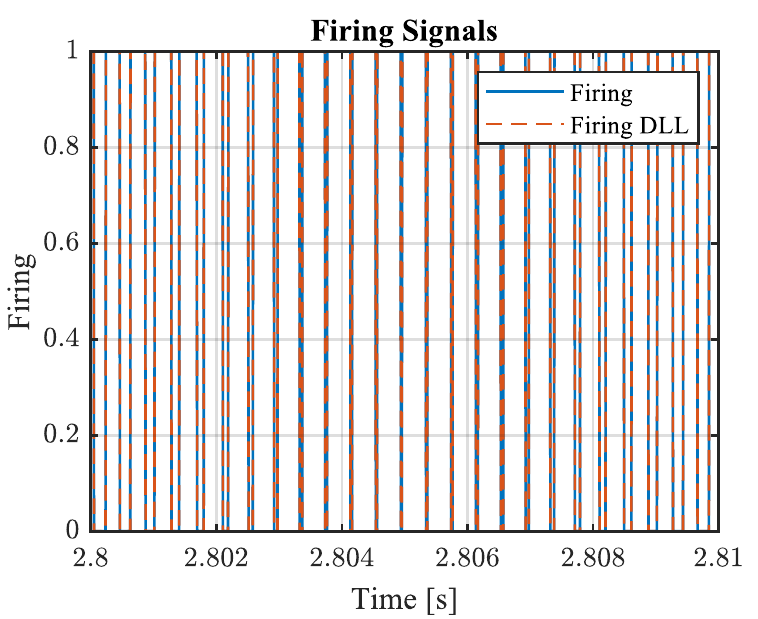}
        \caption{Firing - Small time step}
        \label{fig:fi-fast}
    \end{subfigure}
    \begin{subfigure}[b]{0.32\textwidth}
        \includegraphics[width=\linewidth]{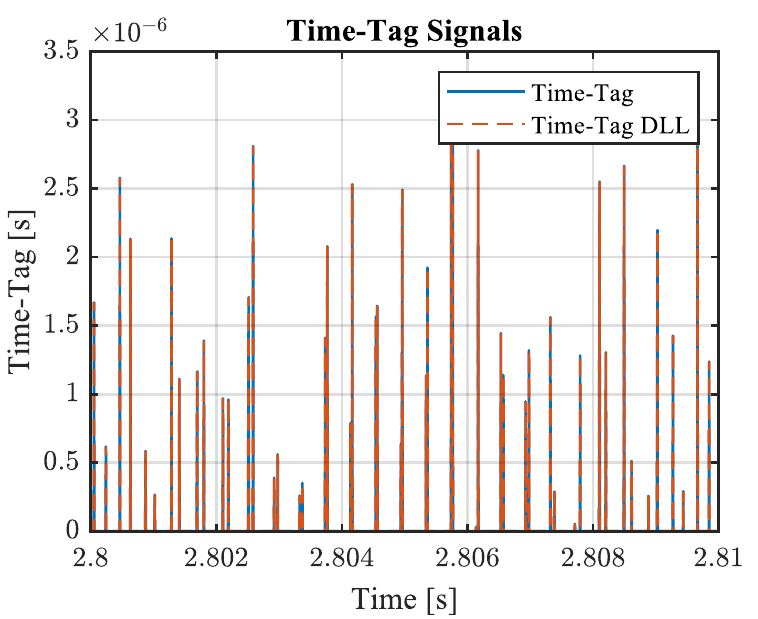}
        \caption{$\text{Time-Tag}$ - Small time step}
        \label{fig:tt-fast}
    \end{subfigure}

    \caption{Interpolated firing block and DLL version for different time steps and signal frequencies}
    \label{fig:confronto-segnali}
\end{figure*}

Initial tests aimed to determine whether integrating signal generation through the DLL into the black-box model could enhance simulation accuracy. These tests checked if the DLL could accurately replicate the Interpolated Firing Signals generator across different simulation time steps. They also evaluated the impact of DLL-based implementations on simulation speed by measuring CPU time. Since the C code was auto-generated from Simulink, it's important to note that the simulation time in Simulink does not directly match the timing of the DLLs. The same logic was applied consistently across the entire switching module. Lastly, tests were carried out to assess whether the DLL influences the OEM's harmonic model.

\subsection{Simple Interpolated Firing block comparison}
In the first test, shown in \figref{fig:confronto-segnali}, the EMT's software (in this case, PSCAD) compared the Interpolated firing block generator with its equivalent DLL version under two conditions, as shown in \tabref{tab:firing}.

\begin{table}[h]
\small
\caption{Firing Block Comparison Cases}
    \centering
    \begin{tabular}{cccc}
      \toprule
      Case & Time Step & Modulant Frequency & Carrier Frequency \\
      \midrule
        1 & $0.01s$ & $1Hz$ & $2Hz$ \\
        2 & $3.125 \mu s$  & $50Hz$ & $2500Hz$ \\
        \bottomrule
    \end{tabular}
    \label{tab:firing}
\end{table}

In particular, it is recognized that the results of the in-house firing signal generator and the DLL version match for both firing signal patterns and $\text{Time-Tag}$ magnitudes. However, it must be highlighted that for enhancing DLL's fidelity (in PSCAD), it was necessary either to use a modulant signal generated through a \textit{Voltage Controlled Oscillator} by generating a pure signal, or to use a voltage measured directly from a circuit, and to implement the specialized \textit{DSDyn} block that forces the voltage (usually processed in the \textit{DSout} routine) to be done in the \textit{DSDyn}. This is because some EMT software separates the pure signals and mathematical calculations from the electrical solutions into two distinct routines (\textit{DSDyn} and \textit{DSout}), which are executed sequentially by one time step \cite{Manitoba-HVDCResearchCentre2010EMTDC-TransientSimulation}. Therefore, for applications that necessitate high-speed resolution, this can significantly affect the results, specifically in this case by switching the IGBTs one time-step delayed.

For what concerns the simulation speed with DLLs, the CPU time decreased from $12,000 ms$ to $2,985 ms$ over a total duration of $5 s$. This improvement resulted from removing redundant states, applying algebraic operations instead of states when possible, see example in  \eqref{eqn: sindelay}, and calling the interpolation function \eqref{eqn:tt} only when necessary.

\subsection{Full Switching block comparison with synthetic signals}

\begin{figure}[h!]
    \centering
    \begin{subfigure}[b]{0.8\linewidth}
        \centering
        \includegraphics[width=\linewidth]{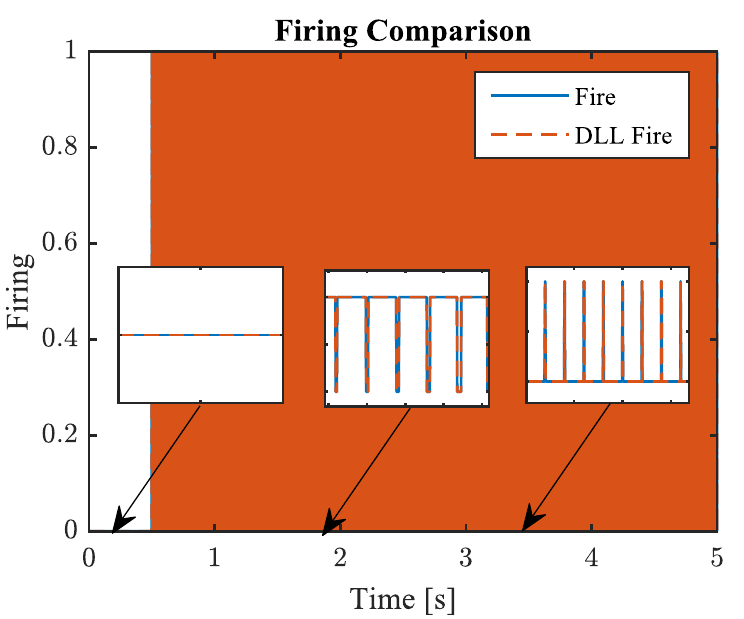}
        \caption{Fire Full switching with synthetic inputs}
        \label{fig:plotA}
    \end{subfigure}
    \hfill
    \begin{subfigure}[b]{0.8\linewidth}
        \centering
        \includegraphics[width=\linewidth]{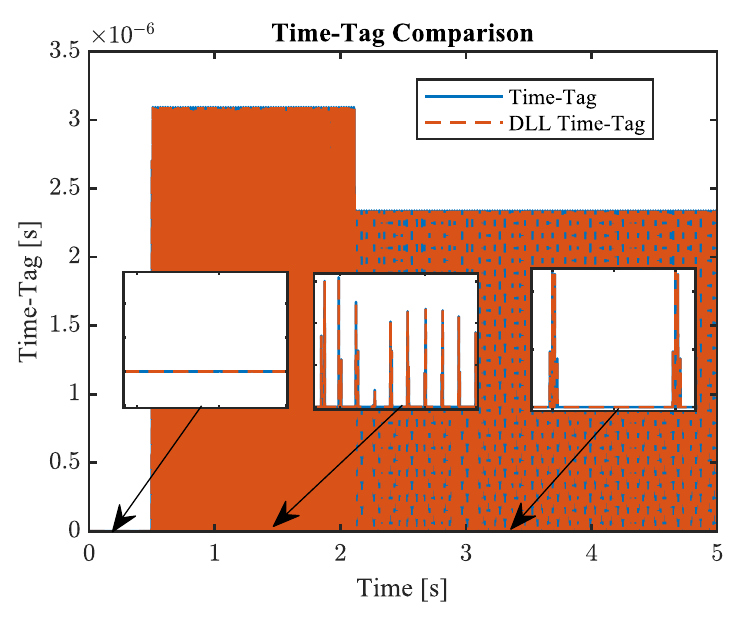}
        \caption{$\text{Time-Tag}$ full switching with synthetic inputs}
        \label{fig:plotB}
    \end{subfigure}
    \caption{Full Switching module comparison with synthetic inputs}
    \label{fig:fullPWM}
\end{figure}

The next step involved translating the overall switching logic from the original OEM's PSCAD model to Simulink, which was then converted into auto-generated C code, resulting in a DLL. The entire switching logic is composed of various patterns, including blocking, regular operation, DC chopper activation, etc. The logic was validated by inputting synthetic signals to isolate the three different behaviors, thereby debugging the overall code until the DLL matched the original behavior. The results are shown in \figref{fig:fullPWM}.

As with the previous case, you can see how, with synthetic signals, the DLL and the original OEM's version match. However, it is crucial to note that the carrier signal must be generated in the DLL differently from the case shown in \figref{fig:confronto-segnali}. 

This requires creating signals, such as the saw-tooth function, using fundamental Simulink blocks or pure code. It is also important to note that these signals must not depend on an internal clock; therefore, either a discrete counter is used internally in the DLL, or these signals can be considered time-dependent on the final simulation clock, which is treated as an input to the overall DLL along with the other signals.
Additionally, whenever periodical signals are generated through pure code, additional attention should be paid to the machine error correlated with functions such as the $mod$ (that creates the signal periodicity). In this case, additional care should be taken to round explicitly all the signals that have a discrete form; otherwise, the associated risk is the generation of relatively random signals that can change one time-step earlier or later, thus affecting the overall switching as previously mentioned.

\subsubsection{Final test on complete OEM model}
\begin{figure}[t!]
    \centering
    \includegraphics[width=1\linewidth]{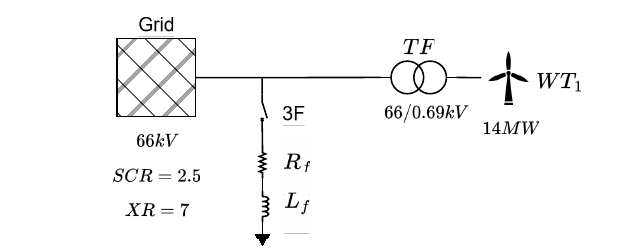}
    \caption{Scheme}
    \label{fig:scheme}
\end{figure}

The final test was performed by substituting the DLL into the full OEM's model and comparing the original behavior. The OEM's WT is connected to an equivalent Thevenin circuit as shown in  \figref{fig:scheme}. 

\begin{figure}
    \centering
    \begin{subfigure}[b]{0.8\linewidth}
        \centering
        \includegraphics[width=\linewidth]{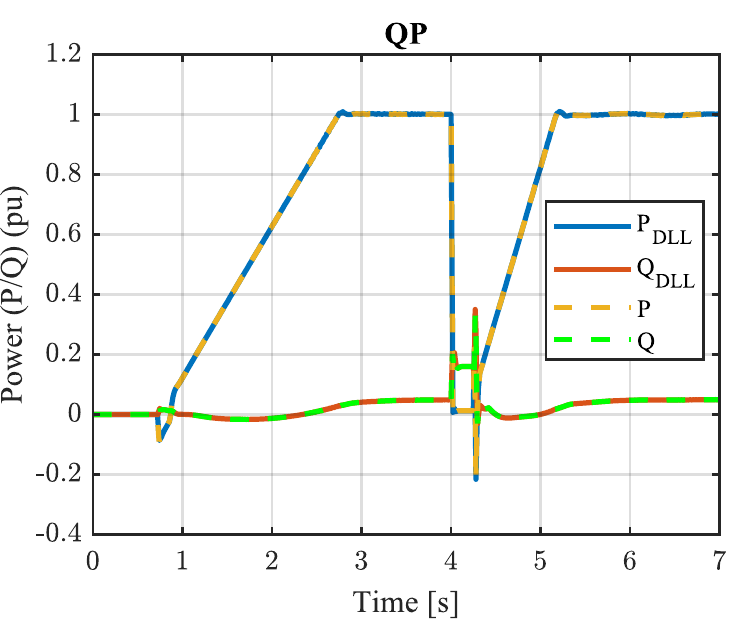}
        \caption{Active/reactive power}
        \label{fig:plotA}
    \end{subfigure}
    \hfill
    \begin{subfigure}[b]{0.8\linewidth}
        \centering
        \includegraphics[width=\linewidth]{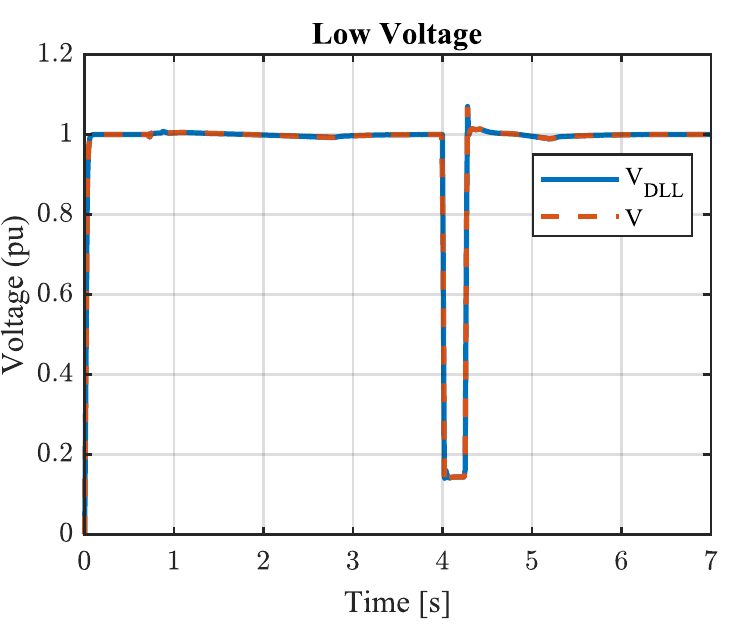}
        \caption{Voltage}
        \label{fig:plotB}
    \end{subfigure}
    \caption{Comparison of two plots side by side in a single column.}
    \label{fig:two_side_by_side}
\end{figure}

After reaching the steady state around $2.5 s$, a three-phase fault occurs at $4 s$ with a consequent fault-ride-through. In this setup, the DLL no longer operates with synthetic signals; therefore, there are no longer pure $50 \text{Hz}$ sinewave modulated signals or clear separation between the three phases, as depicted in \figref{fig:fullPWM}. 
By examining the Active/Reactive power and voltage behaviour in \figref{fig:two_side_by_side}, it is possible to appreciate that the DLL version is indistinguishable from the version without DLL.

\begin{figure*}[ht!]
    \centering
    \begin{subfigure}[b]{0.19\textwidth}
        \centering
        \includegraphics[width=\linewidth]{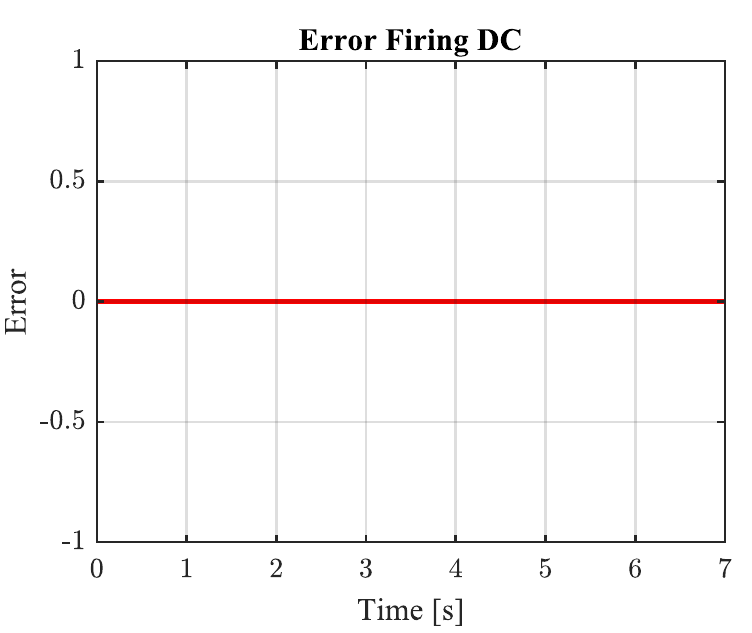}
        \caption{Error DC firing}
        \label{fig:plot1}
    \end{subfigure}
    \begin{subfigure}[b]{0.19\textwidth}
        \centering
        \includegraphics[width=\linewidth]{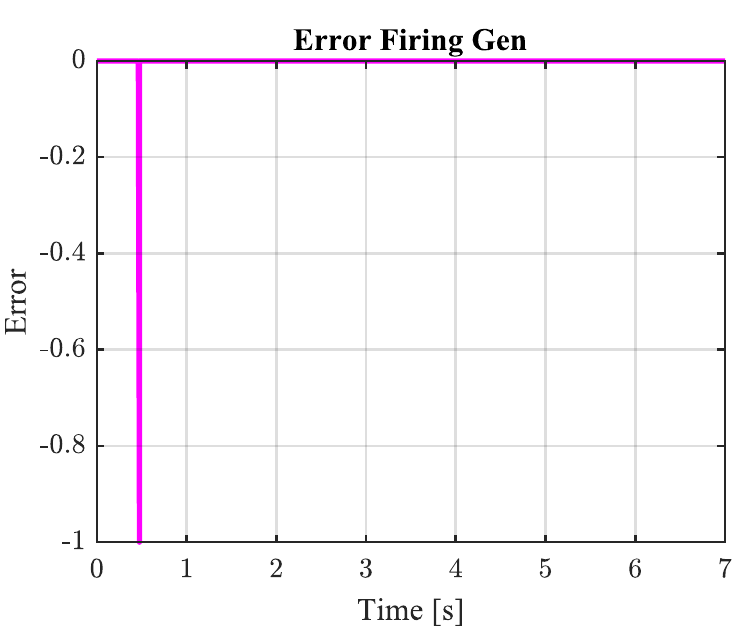}
        \caption{Error Gen-side firing}
        \label{fig:plot2}
    \end{subfigure}
    \begin{subfigure}[b]{0.19\textwidth}
        \centering
        \includegraphics[width=\linewidth]{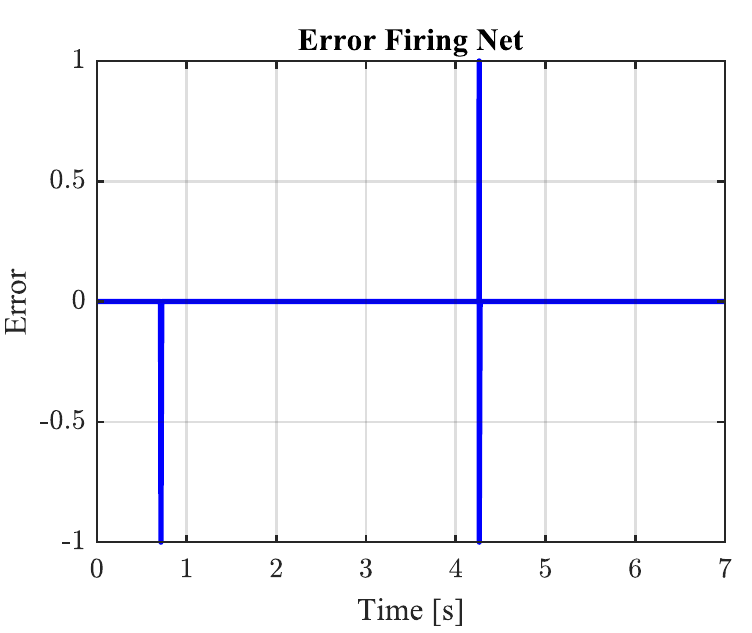}
        \caption{Error Net-side firing}
        \label{fig:plot3}
    \end{subfigure}
    \begin{subfigure}[b]{0.19\textwidth}
        \centering
        \includegraphics[width=\linewidth]{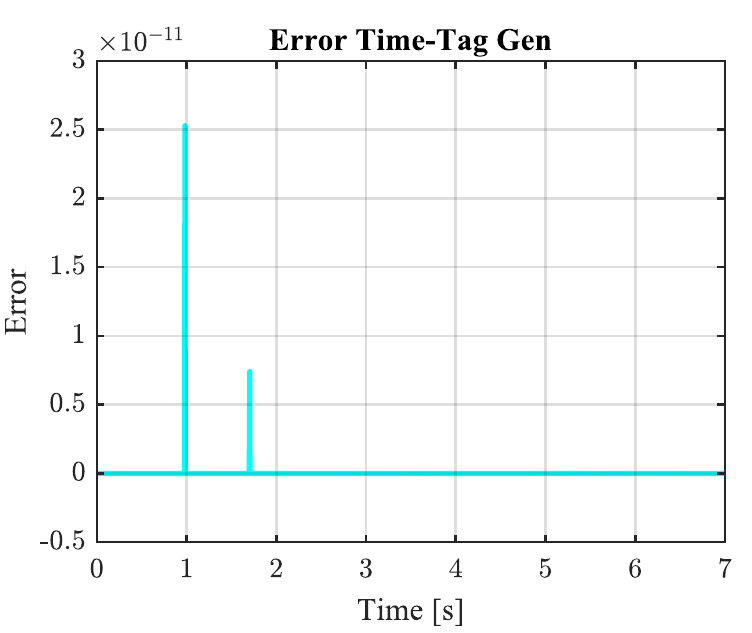}
        \caption{Error Gen $\text{Time-Tag}$}
        \label{fig:plot4}
    \end{subfigure}
    \begin{subfigure}[b]{0.19\textwidth}
        \centering
        \includegraphics[width=\linewidth]{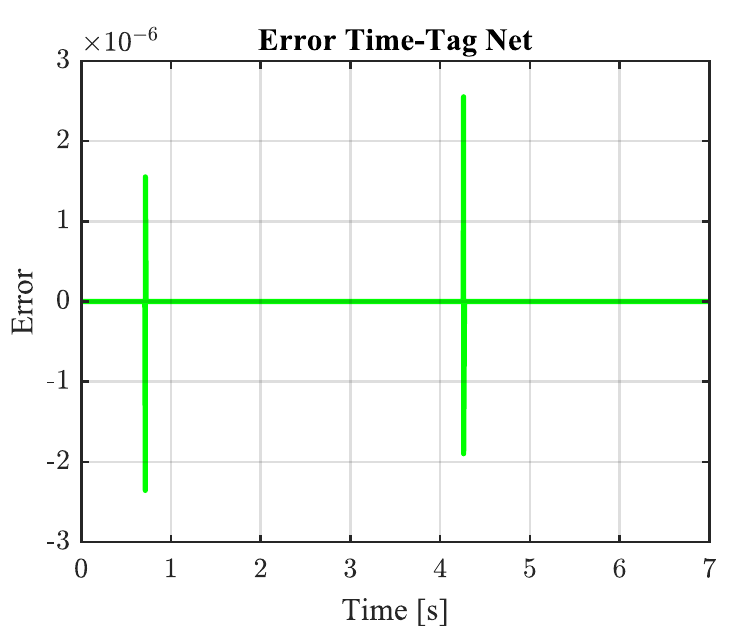}
        \caption{Error Net $\text{Time-Tag}$}
        \label{fig:plot5}
    \end{subfigure}
    
    \begin{subfigure}[b]{0.19\textwidth}
        \centering
        \includegraphics[width=\linewidth]{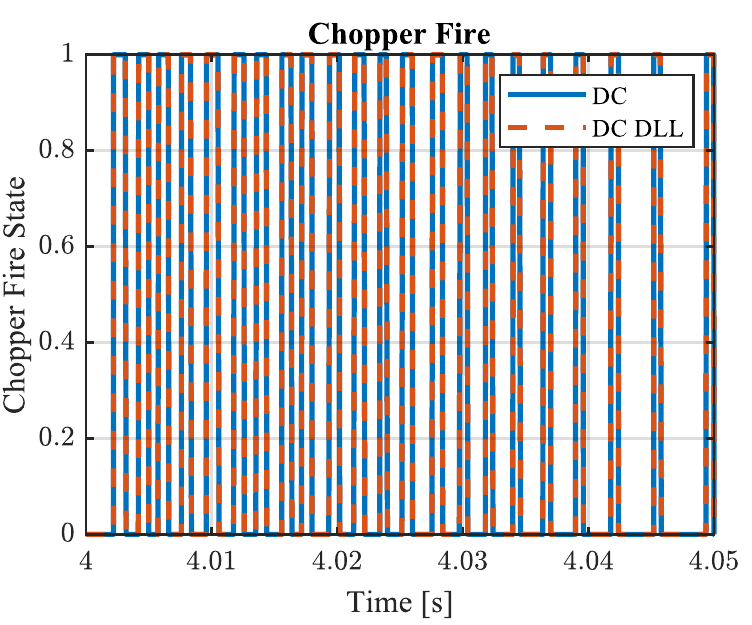}
        \caption{Chopper fire}
        \label{fig:plot6}
    \end{subfigure}
    \begin{subfigure}[b]{0.19\textwidth}
        \centering
        \includegraphics[width=\linewidth]{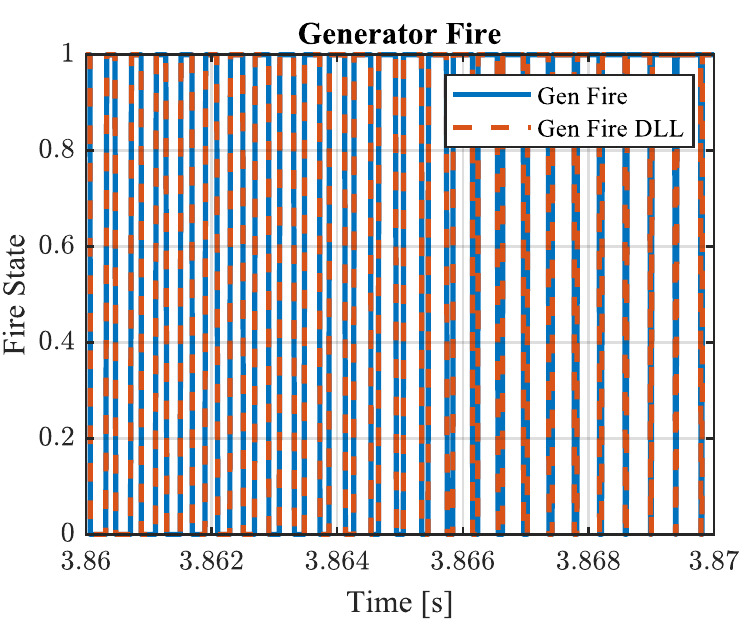}
        \caption{Gen fire}
        \label{fig:plot7}
    \end{subfigure}
    \begin{subfigure}[b]{0.19\textwidth}
        \centering
        \includegraphics[width=\linewidth]{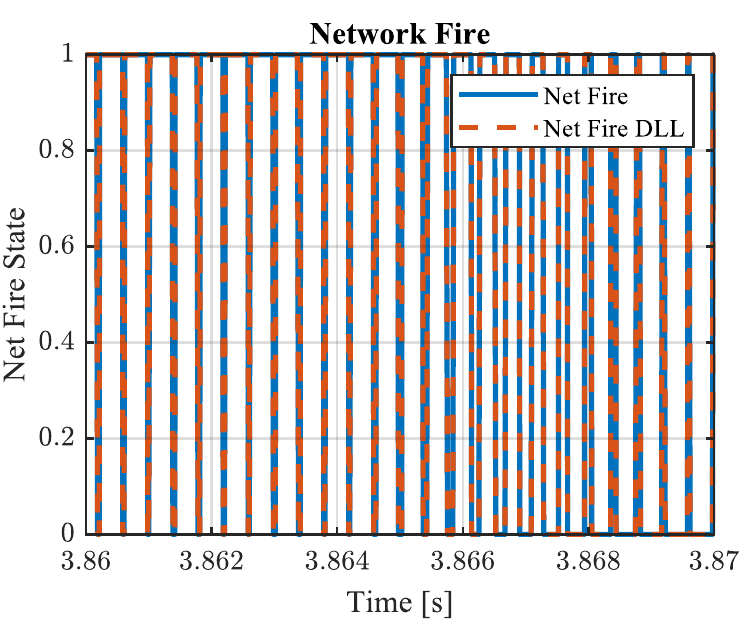}
        \caption{Net fire}
        \label{fig:plot8}
    \end{subfigure}
    \begin{subfigure}[b]{0.19\textwidth}
        \centering
        \includegraphics[width=\linewidth]{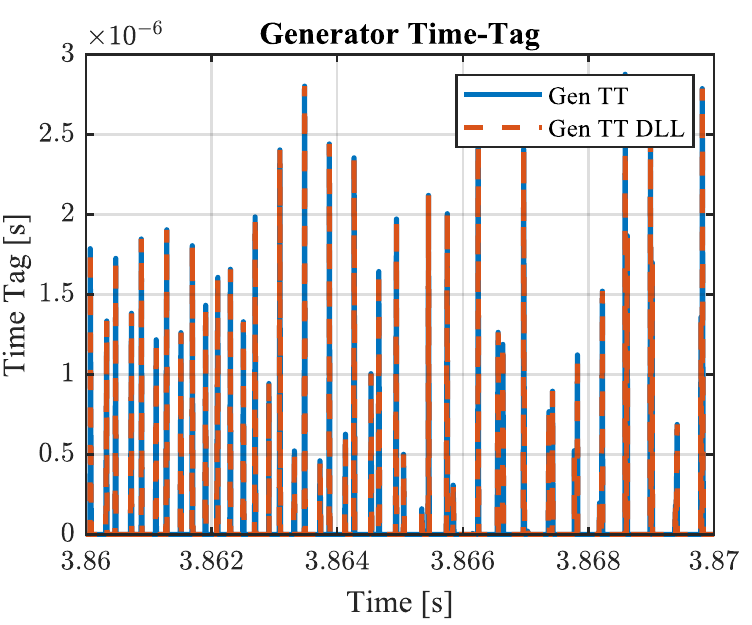}
        \caption{Gen $\text{Time-Tag}$}
        \label{fig:plot9}
    \end{subfigure}
    \begin{subfigure}[b]{0.19\textwidth}
        \centering
        \includegraphics[width=\linewidth]{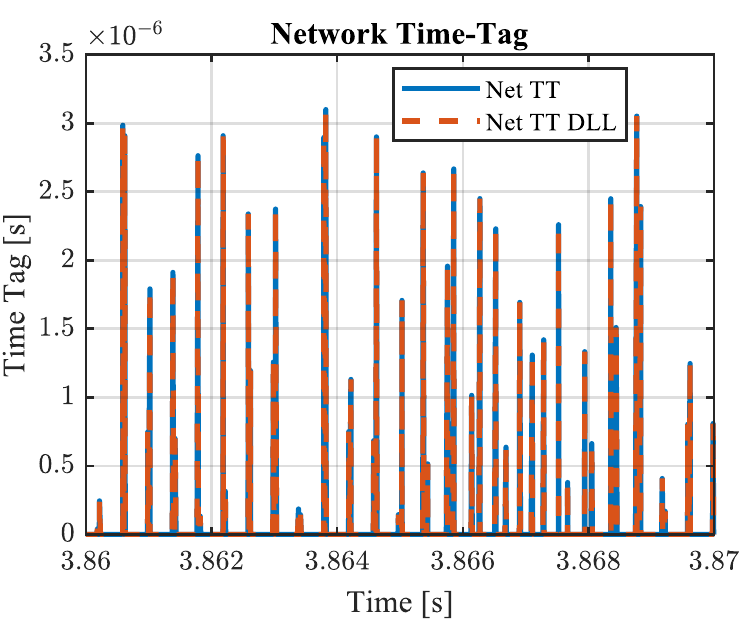}
        \caption{Net $\text{Time-Tag}$}
        \label{fig:plot10}
    \end{subfigure}
    
    \caption{Comparison between the expected original signals and the DLL in the final benchmark}
    \label{fig:ten_subplots}
\end{figure*}

To assess whether the DLL-based version differs from the native EMT one, it is useful to analyze the DLL's output signals directly. As shown in 
\figref{fig:ten_subplots}, the error plots reveal some singular events that cause minor errors. These usually occur during mode switches—such as switching from blocking to regular operation or during a fault/over-current scenario. The multiple errors observed at a single time-step shift are caused by the network during current limiting (following a fault at $4s$), likely due to feedback loops in the current limiting logic modeled with Simulink, which uses auto-generated C code. 

Since memory blocks are necessary to prevent algebraic loops, this setup introduces a one-time-step delay. The current limiting operation is generally not continuous; instead, it may switch rapidly as the current crosses the threshold. Therefore, the error is not visible as a single event but is repeated multiple times during a fault condition.
Additionally, since the current measurement is performed in the $DSout$ routine, which is delayed by one time step, this explains the repeated one-time-step shift observed during the fault condition in \figref{fig:ten_subplots} in the error plots. 

To sum up, the full switching module in DLL can be regarded as equivalent to the in-house EMT's model version. Although there are minor differences at switching instants, careful analysis shows that the performance is indistinguishable in terms of active and reactive power during both normal and FRT routines. 

\subsection{TSO's Perspective}
The previous sections demonstrated that the DLL version of the switching model can be compared with the original EMT version, starting with simple synthetic signals and progressing to FRT situations. The tests were conducted at the nominal simulation time step; however, this may not be sufficient from the TSO's perspective, since grid codes require shared models to be compatible and valid across multiple time-step applications \cite{SimulationEirGrid}. This section analyzes how the same unaltered DLL-based version would perform when the simulation time step is increased to 10 microseconds (three times larger). 
From \figref{fig:ten_subplots}, it is evident that, based on the PQ requirements—which are likely the main concern for the TSOs—the DLL version is almost identical to that at 10us. Examining the error pattern in the switching signals, it is clear that, from the grid side (the TSOs' area of interest), the error patterns are nearly identical.

From the OEM perspective, the version at the highest time step does not completely overlap with the DC chopper's switching pattern. At the generator, it differs significantly from the nominal time-step case. 
Therefore, if the TSO requires a different, larger time step for the PQ signals in the steady and FRT cases, they remain compliant. The trade-off is that some internal variables, which may not align with the TSO's interests, will not exactly match the nominal case. The importance of success, then, depends on the OEMs' judgment in determining whether the outputs are sufficiently valid based on their physical models and standards.

\begin{figure*}[ht!]
    \centering
    \begin{subfigure}[b]{0.32\textwidth}
        \centering
        \includegraphics[width=\linewidth]{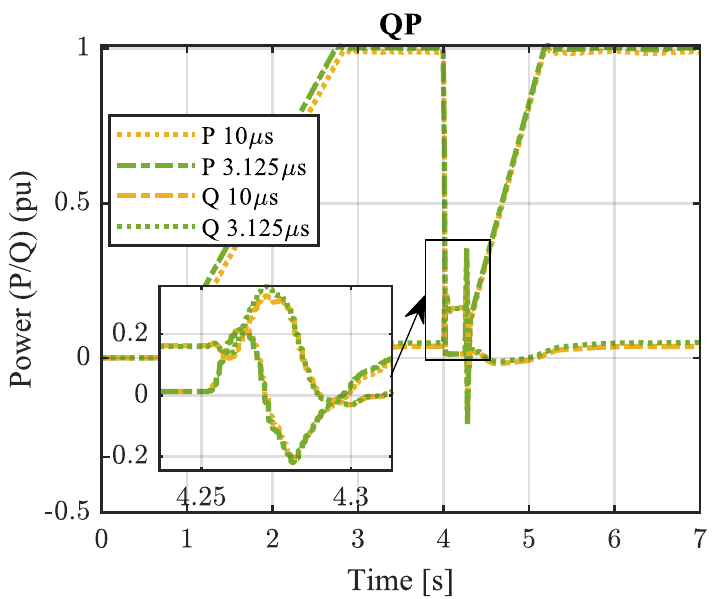}
        \caption{Active/reactive power larger time step}
        \label{fig:2plot1}
    \end{subfigure}
    \hfill
    \begin{subfigure}[b]{0.32\textwidth}
        \centering
        \includegraphics[width=\linewidth]{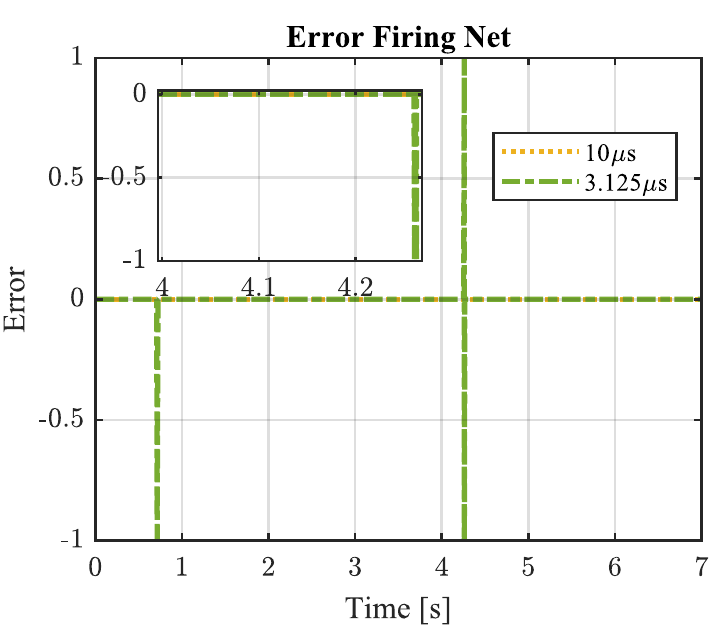}
        \caption{Error Net-side firing larger time step}
        \label{fig:2plot2}
    \end{subfigure}
    \hfill
    \begin{subfigure}[b]{0.32\textwidth}
        \centering
        \includegraphics[width=\linewidth]{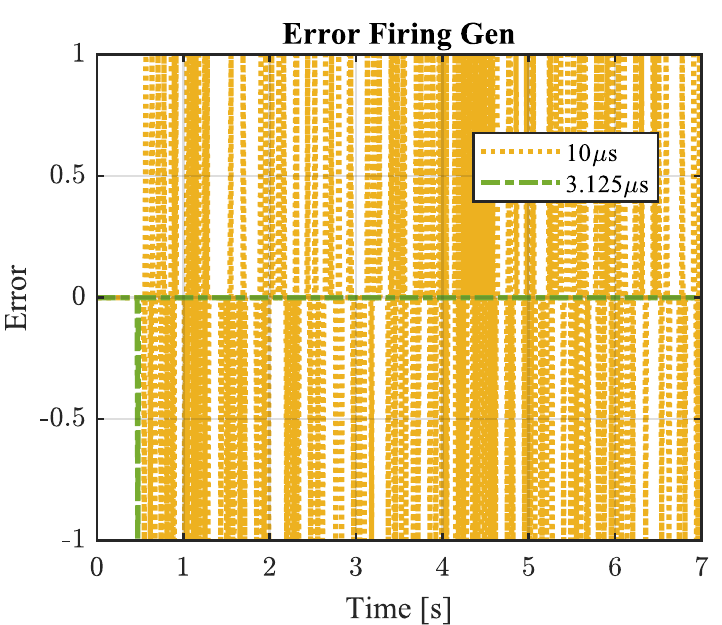}
        \caption{Error Gen-side firing larger time step}
        \label{fig:2plot3}
    \end{subfigure}
    
    \begin{subfigure}[b]{0.32\textwidth}
        \centering
        \includegraphics[width=\linewidth]{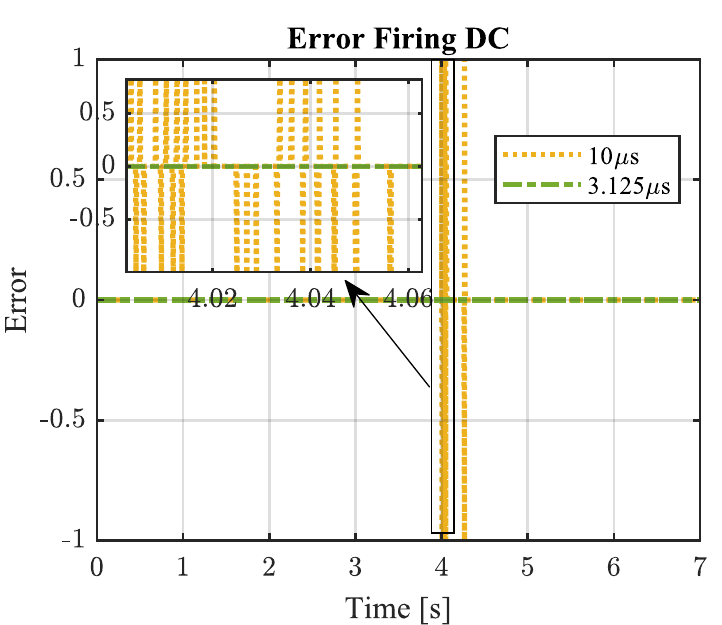}
        \caption{Chopper fire larger time step}
        \label{fig:2plot4}
    \end{subfigure}
    \hfill
    \begin{subfigure}[b]{0.32\textwidth}
        \centering
        \includegraphics[width=\linewidth]{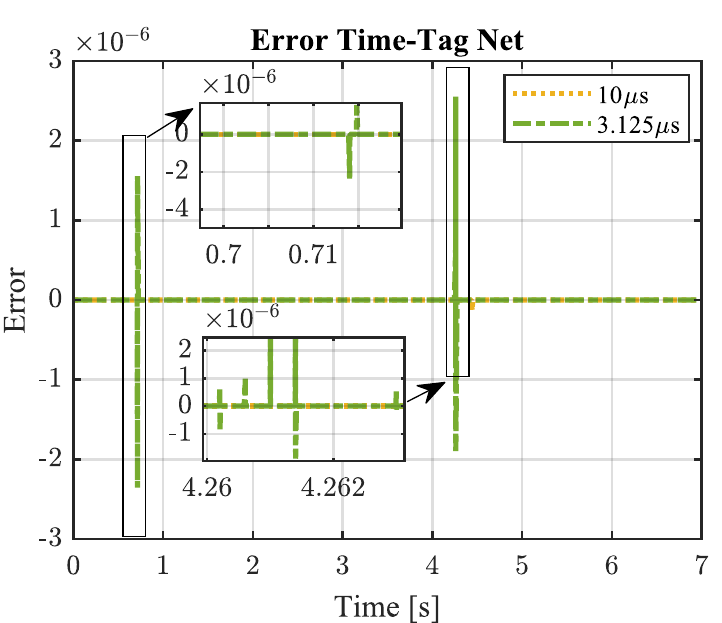}
        \caption{Error Net $\text{Time-Tag}$ larger time step}
        \label{fig:2plot5}
    \end{subfigure}
    \hfill
    \begin{subfigure}[b]{0.32\textwidth}
        \centering
        \includegraphics[width=\linewidth]{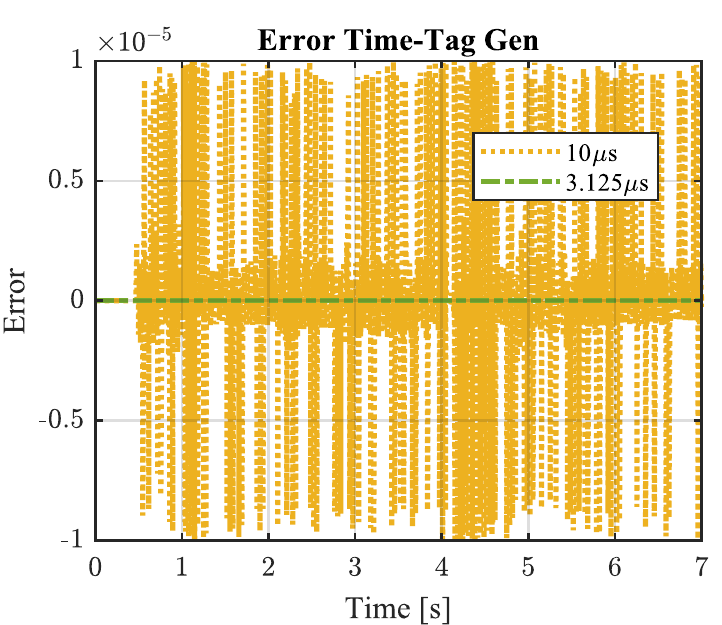}
        \caption{Error Gen $\text{Time-Tag}$ larger time step}
        \label{fig:2plot6}
    \end{subfigure}
    
    \caption{Comparison between the expected original signals and the DLL in the final benchmark at larger simulation's time step}
    \label{fig:six_subplots}
\end{figure*}

\subsection{DLL effects on the harmonic models}

\begin{figure}[t!]
    \centering
    \includegraphics[width=0.8\linewidth]{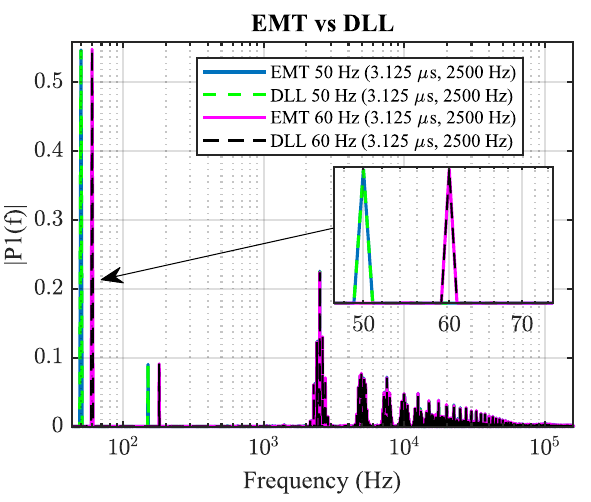}
    \caption{Harmonic content PWM}
    \label{fig: harmonic}
\end{figure}

To verify whether the DLL may affect the harmonic models, the harmonic spectrum of the PWM switching signal for a single phase was first analyzed, both with and without the DLL. It is concluded that, in terms of harmonic content introduced by the switching block, the DLL matches well the original EMT version as shown in \figref{fig: harmonic}.

The whole $\text{Time-Tag}$ logic aids understanding of when a signal actually transitioned by providing an additional scalar value in parallel to the firing signal, as shown in \figref{fig:difference}. Therefore, if only the firing signal is used to study the harmonic content, this may be biased, since it does not represent the actual switching signal. Then it may be argued that the harmonic analysis performed in \figref{fig: harmonic}, directly on the switching signals with and without DLL, is biased. 

To eliminate any doubts about the coincidence of the harmonic models, the analysis was also performed on the output terminal current and voltages. This is because these results are from the second part of the EMTDC process (\textit{DSOut}) \cite{Manitoba-HVDCResearchCentre2010EMTDC-TransientSimulation}, where the $\text{Time-Tag}$ correction is already included in the switching, thus in the converter's outputs. 

\begin{figure}
    \centering
    \begin{subfigure}[b]{0.8\linewidth}
        \centering
        \includegraphics[width=\linewidth]{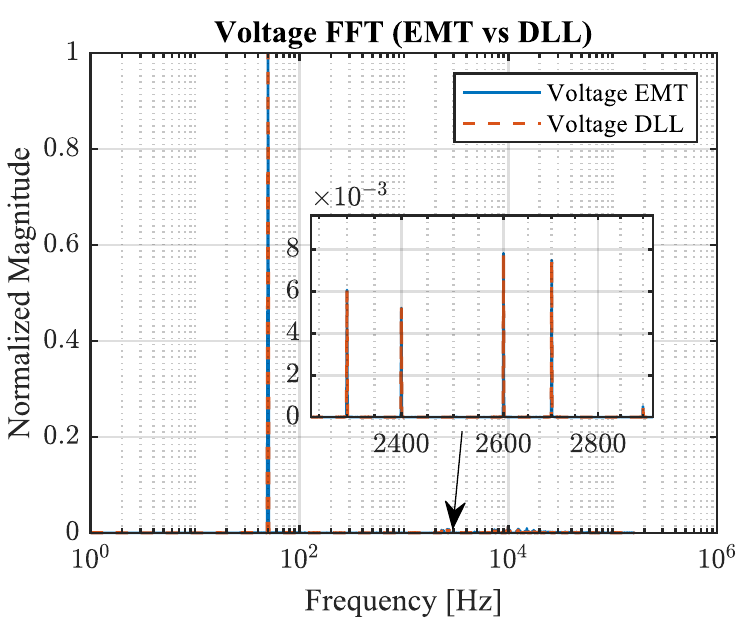}
        \caption{Voltage FFT}
        \label{fig:plotA}
    \end{subfigure}
    \hfill
    \begin{subfigure}[b]{0.8\linewidth}
        \centering
        \includegraphics[width=\linewidth]{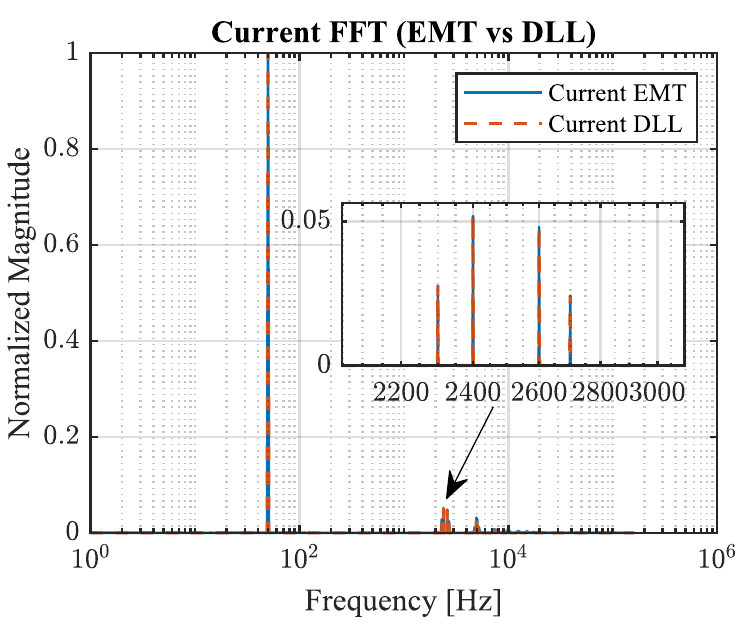}
        \caption{Current FFT}
        \label{fig:plotB}
    \end{subfigure}
    \caption{Voltage and Current FFT with and without DLL}
    \label{fig: FFT_C_V}
\end{figure}

The harmonic analysis performed on the output currents and voltages visible in \figref{fig: FFT_C_V} shows that the DLL and original EMT version match well.

Therefore, it can be concluded that the DLL version does not differ from the original representation in PSCAD, even when considering the harmonic model perspective. This supports the feasibility of substituting the DLL switching module with the emulated interpolation.  

\section{Discussion and Conclusion}
The modeling and sharing of EMT models are becoming increasingly important in power systems featuring IBR-based devices, where multiple actors interact through fast dynamics. The need for each OEM to protect the internal structure of their devices adds a layer of complexity when identifying potential issues. For these reasons, sharing parts of models via DLLs—especially those representing actual device firmware—is becoming standard practice. This approach facilitates studies involving multiple stakeholders and complex components and improves both system accuracy and model security.
In this context, the inclusion of switching generation signals in DLLs is critical to further enhance simulation fidelity. However, implementing mechanisms such as blocking, alongside switching, is not straightforward: control parts involve interpolation mechanisms for high-frequency signals that transition between time steps. This study has focused specifically on implementing these mechanisms within the DLL, assessing their feasibility and impact on simulation speed.
The methodology for incorporating switching behavior in the DLL first involved reviewing the interpolation mechanism and accurately translating it into Simulink/C Code, so that all interpolation-based components could be reproduced in code. Initial validation was carried out by testing elementary blocks—such as the Interpolated Firing Signal Generator—and then assembling the complete switching module for further verification using both synthetic and realistic signals. Ultimately, we integrated the DLL into the full OEM model and tested it under a fault ride-through scenario.
The various tests using the nominal time step accurately captured the EMT-based versions from any perspective, confirming the possibility of including interpolation-based elements in DLLs. There are minor discrepancies during mode transitions where the one-time-step delay plays a significant role, yet it is irrelevant in terms of output fidelity. 
Regarding DLL speed, we observed that the number of states in the DLL affects its CPU time. While some actions made the simulation faster in the C version, there was no clear correlation between actions that sped up the source code and those that sped up the DLL. This makes the speed-up process iterative. The research then proposed some actions, such as computing the $\text{Time-Tag}$ only when there was a switching delay, which had a clear effect on the output. 
Nevertheless, some limitations must be recognized. The accuracy of DLL-based models can be impacted by delays from the EMT-DLL interface, especially when the DLL logic depends heavily on input signals from measurements (e.g., current, DC voltage) and is processed at fixed simulation time steps. These issues can become more significant with larger time steps or when additional memory blocks are needed for stability, leading to slight one-step errors during sudden changes in external conditions. Careful attention is therefore essential in interface design and signal management when implementing DLLs across different EMT platform environments.
Although the simulation's time step influences the DLL's accuracy, this is mainly due to the OEM's internal requirements. Based on the outputs required by TSO, it was evident that the DLL version would remain compliant even with a threefold increase in the time step relative to the nominal case. 
Building on this experience, future work should expand and validate the method across various software and hardware platforms, and develop and test advanced strategies and algorithms to enhance accuracy, speed, and compliance with industrial and grid standards. These steps will further strengthen the use of DLL-based black-box models for the validation of offshore wind systems, while promoting cooperation between OEMs and grid operators to enable more accurate and transferable simulation studies in both industrial and regulatory contexts.
In summary, the presented methodology enables the creation of high-fidelity black-box models in DLL format, offering a practical solution for sharing and deeply analyzing complex devices through EMT simulations, with clear avenues for further refinement and future research.

\section{Acknowledgements}
This work is supported by the European Union as part of ADOreD project funded by the Horizon Europe MSCA programme (\href{https://www.msca-adored.eu/}{HORIZON-MSCA-2021-DN, Grant agreement 101073554})

\section{Legal Disclaimer}
Figures and values presented in this paper should not be used to judge the performance of Siemens Gamesa Renewable Energy technology as they are solely presented for demonstration purpose. Any opinions or analysis contained in this paper are the opinions of the authors and not necessarily the same as those of Siemens Gamesa Renewable Energy.

\bibliographystyle{unsrt}  
\bibliography{references}

@article{Deng2024AStudy,
    title = {{A Novel Approach for Harmonic Assessment of Power Systems With Large Penetration of IBRs - A U.K. Case Study}},
    year = {2024},
    journal = {IEEE Transactions on Power Delivery},
    author = {Deng, Zhida and Todeschini, Grazia},
    number = {1},
    month = {2},
    pages = {455--466},
    volume = {39},
    publisher = {Institute of Electrical and Electronics Engineers Inc.},
    doi = {10.1109/TPWRD.2023.3265902},
    issn = {19374208}
}

@article{Jahn2022AnProtection,
    title = {{An Architecture for a Multi-Vendor VSC-HVDC Station with Partially Open Control and Protection}},
    year = {2022},
    journal = {IEEE Access},
    author = {Jahn, Ilka and Nahalparvari, Mehrdad and Hirsching, Carolin and Hoffmann, Melanie and Dullmann, Patrick and Loku, Fisnik and Agbemuko, Adedotun and Chaffey, Geraint and Prieto-Araujo, Eduardo and Norrga, Staffan},
    pages = {13555--13569},
    volume = {10},
    publisher = {Institute of Electrical and Electronics Engineers Inc.},
    doi = {10.1109/ACCESS.2022.3146782},
    issn = {21693536}
}

@misc{CSEECIGRE,
    title = {{CSE 028 - CIGRE Science {\&} Engineering | eCIGRE}},
    url = {https://www.e-cigre.org/publications/detail/cse028-cse-028.html}
}

@techreport{2025DefinitionPUBLIC,
    title = {{Definition of a standard process for interaction studies with EMT simulation in multi-vendor projects 1 Definition of a standard process for interaction studies with EMT simulation in multi-vendor projects PUBLIC}},
    year = {2025},
    month = {3},
    institution = {InterOPERA}
}

@article{Manitoba-HVDCResearchCentre2010EMTDC-TransientSimulation,
    title = {{EMTDC-Transient Analysis for PSCAD Power System Simulation}},
    year = {2010},
    author = {{Manitoba-HVDC Research Centre}}
}

@article{Gharehpetian2024FutureSolutions,
    title = {{Future Power System Elements, Challenges, and Solutions}},
    year = {2024},
    journal = {Future Power System Elements, Challenges, and Solutions},
    author = {Gharehpetian, Gevork B. and Zolfaghari, Mahdi and Bayati, Navid},
    month = {1},
    pages = {1--309},
    publisher = {Elsevier},
    isbn = {9780443140914},
    doi = {10.1016/C2022-0-03146-9}
}

@misc{GitHubRte-france/PSCAD-import-tool-for-IEEE-CIGRE-DLLs,
    title = {{GitHub - rte-france/PSCAD-import-tool-for-IEEE-CIGRE-DLLs}},
    url = {https://github.com/rte-france/PSCAD-import-tool-for-IEEE-CIGRE-DLLs}
}

@misc{GuidelinesECIGRE,
    title = {{Guidelines for use of real-code in EMT models for HVDC, FACTs and inverter based generators in power systems analysis - Technical Brochures | eCIGRE}},
    url = {https://www.e-cigre.org/publications/detail/958-guidelines-for-use-of-real-code-in-emt-models-for-hvdc-facts-and-inverter-based-generators-in-power-systems-analysis.html}
}

@misc{IECIEC,
    title = {{IEC 61400-27-1:2020 | IEC}},
    url = {https://webstore.iec.ch/en/publication/32564}
}

@article{Beddard2016ImprovedConverters,
    title = {{Improved accuracy average value models of modular multilevel converters}},
    year = {2016},
    journal = {IEEE Transactions on Power Delivery},
    author = {Beddard, A. and Sheridan, C. E. and Barnes, M. and Green, T. C.},
    number = {5},
    month = {10},
    pages = {2260--2269},
    volume = {31},
    publisher = {Institute of Electrical and Electronics Engineers Inc.},
    doi = {10.1109/TPWRD.2016.2535410},
    issn = {08858977}
}

@article{Marthi2023InterpolationInverters,
    title = {{Interpolation Methods to Enable Fast and Accurate EMT Simulation of PV Inverters}},
    year = {2023},
    journal = {2023 IEEE 24th Workshop on Control and Modeling for Power Electronics, COMPEL 2023},
    author = {Marthi, Phani R.V. and Debnath, Suman and Choi, Jongchan},
    publisher = {Institute of Electrical and Electronics Engineers Inc.},
    isbn = {9798350316186},
    doi = {10.1109/COMPEL52896.2023.10221104}
}

@misc{SimulationEirGrid,
    title = {{Simulation and Modelling | Grid Codes and Compliance | EirGrid}},
    url = {https://www.eirgrid.ie/grid/grid-codes-and-compliance-overview/simulation-studies-and-modelling-requirements}
}

@article{Watzlawick1988Ultra-Solutions:Successfully,
    title = {{Ultra-Solutions: How to Fail Most Successfully}},
    year = {1988},
    author = {Watzlawick, Paul.},
    pages = {112},
    publisher = {W.W. Norton {\&} Company, Incorporated},
    isbn = {0393333760}
}

\end{document}